# Two-scale momentum theory for very large wind farms

**Takafumi Nishino**

Centre for Offshore Renewable Energy Engineering, Cranfield University, Cranfield, Bedfordshire MK43 0AL, United Kingdom

E-mail: t.nishino@cranfield.ac.uk

**Abstract**. A new theoretical approach is proposed to predict a practical upper limit to the efficiency of a very large wind farm. The new theory suggests that the efficiency of ideal turbines in an ideal very large wind farm depends primarily on a non-dimensional parameter $\lambda/C_{f0}$, where $\lambda$ is the ratio of the rotor swept area to the land area (for each turbine) and $C_{f0}$ is a natural friction coefficient observed before constructing the farm. When $\lambda/C_{f0}$ approaches to zero, the new theory goes back to the classical actuator disc theory, yielding the well-known Betz limit. When $\lambda/C_{f0}$ increases to a large value, the maximum power coefficient of each turbine reduces whilst a normalised power density of the farm increases asymptotically to an upper limit. A CFD analysis of an infinitely large wind farm with 'aligned' and 'displaced' array configurations is also presented to validate a key assumption used in the new theory.

## 1. Introduction
Evaluating the efficiency of a wind farm is not a trivial task. Even without considering any financial factors that affect the overall (or economic) efficiency of a wind farm, aerodynamic efficiency of a number of turbines arrayed as a farm is much more difficult to evaluate compared to that of a single isolated turbine. A major problem here is that we do not have a good 'absolute' (rather than relative) basis of evaluation for farm efficiency, such as the well-known 'Betz limit' [1] for single turbine efficiency. The lack of such an absolute basis of evaluation makes it very difficult to evaluate how good, or not so good, the efficiency of a given/existing wind farm really is.

   In this study, I propose a new theoretical approach to predict (or at least help predict) a practical upper limit to the efficiency of a very large wind farm. Here 'very large' implies that the horizontal extent of the wind farm is at least an order of magnitude larger than the thickness of the atmospheric boundary layer (ABL), which is typically about 1km. For example, Hornsea Project One wind farm (expected to be fully operational in the UK in 2020) can be seen as such a very large wind farm, the airflow through which may approach the so-called 'fully developed' state [2].

## 2. Two-scale momentum theory
A new theoretical approach to predicting a practical upper limit to the efficiency of a very large wind farm is described below, followed by a numerical analysis in Section 3 and discussion in Section 4.

### 2.1. Farm-scale (or ABL-scale) momentum balance
First, we consider a simple streamwise momentum balance for a 'fully developed' ABL over a very large wind farm. For simplicity we assume that the ABL is driven by a constant streamwise pressure gradient and neglect the Coriolis force—the validity of this simplification has been discussed in, e.g. [2]. Here we do not specify the array configuration (aligned, staggered, etc.) of wind turbines but consider that they are arrayed in a horizontally periodic manner, so that a constant horizontal area, $S$, is allocated to each turbine (and therefore the total site area of the farm is $nS$, where $n$ is the number of turbines in the farm). Due to the assumption that the driving force of the ABL is constant and not affected by the farm, we obtain

$$\langle \tau_w \rangle S + T = \tau_{w0} S = \text{const.} \tag{1}$$

where $\langle \tau_w \rangle$ is the 'wall' shear stress (on the ground or sea surface, depending on whether the farm is onshore or offshore) averaged across the area $S$, $T$ is the thrust on one turbine and $\tau_{w0}$ is the 'natural' wall shear stress observed before constructing the farm.

*2.2. Coupling the actuator disc theory with the farm-scale momentum balance*

Next, we try to relate the turbine thrust $T$ in (1) to the characteristics of flow through the so-called 'wind farm layer' near the bottom of the ABL, using the classical actuator disc theory. We consider this 'wind farm layer' as a layer within which the flow is strongly affected by the turbines; hence the thickness or height of this layer, $H_F$, is somewhat larger than the height of each turbine, $H_T$ (although we assume that this layer is still much thinner than the ABL, i.e. $H_T < H_F \ll \delta_{ABL}$). There are many possible ways to define $H_F$ but here we choose a seemingly artificial definition based on the 'natural' or 'undisturbed' ABL profile (instead of the profile disturbed by the farm; the reason for this choice will become clear later). Specifically, $H_F$ is defined such that $U_{F0} = U_{T0}$, where $U_{F0}$ and $U_{T0}$ are the 'undisturbed' wind speed averaged over the farm layer height $H_F$ and over the swept area of a turbine rotor, $A$, respectively, i.e.

$$U_{F0} \equiv \frac{\int_0^{H_F} U_{(K=0)} dz}{H_F} = U_{T0} \equiv \frac{\int U_{(K=0)} dA}{A} \tag{2}$$

where $U_{(K=0)}$ is the undisturbed streamwise velocity observed before constructing the farm; note that this velocity is a function of the vertical coordinate ($z$) only ($z = 0$ corresponds to the ground).

The main difficulty in estimating the turbine thrust $T$ in (1) is that, in a wind farm, the 'upstream' wind speed for each turbine is not fixed but depending on other turbines in the farm; therefore the classical actuator disc theory cannot be directly applied to this case. However, as will be numerically demonstrated later in Section 3, the actuator disc theory still seems to provide a good approximation to the 'maximum' turbine thrust (and also power) for a given 'local' or 'effective' axial induction at the turbine. Here the 'maximum' means the maximum that could be reached by changing the array configuration as well as the turbine design, whilst the 'local' axial induction factor (of a turbine in a periodic array) can be defined as $a^* = (U_F - U_T)/U_F$, where

$$U_F \equiv \frac{\iint \int_0^{H_F} U dz dS}{H_F S} \quad \text{and} \quad U_T \equiv \frac{\int U dA}{A} \tag{3a,b}$$

are the volume-averaged 'farm-layer' wind speed and the face-averaged wind speed over the turbine rotor swept area, respectively. Note that when the value of $S$ increases and approaches to infinity, $U_F$ approaches to $U_{F0}$ and hence $a^*$ will be identical to a common axial induction factor defined for an isolated turbine, $a = (U_{T0} - U_T)/U_{T0}$.

On the basis of the numerical analysis to be presented later, now we can estimate the 'maximum' turbine thrust (for a given set of $U_F$ and $a^*$) by using the classical actuator disc theory with replacing the 'upstream' or 'reference' wind speed[1] with the 'farm-layer' wind speed $U_F$. Since this 'maximum' turbine thrust is likely to give a practical upper limit to the farm efficiency (see Section 4 for further discussion on this point), we take this turbine thrust as the turbine thrust that we are interested in, $T$, which is calculated (by following the classical actuator disc theory; see, e.g. [4] and [5]) as

$$T = \tfrac{1}{2}\rho U_F^2 A \cdot 4\alpha(1-\alpha) \tag{4}$$

where $\rho$ is the density of air and $\alpha = U_T/U_F = (1-a^*)$ is the ratio of the average wind speed over the turbine rotor swept area to that through the farm layer. By substituting (4) into (1) we obtain

$$1 - \frac{\langle \tau_w \rangle}{\tau_{w0}} = \frac{A}{S} \cdot \frac{\tfrac{1}{2}\rho U_{F0}^2}{\tau_{w0}} \cdot \frac{U_F^2}{U_{F0}^2} \cdot 4\alpha(1-\alpha) = \lambda \cdot \frac{1}{C_{f0}} \cdot \beta^2 \cdot 4\alpha(1-\alpha) \tag{5}$$

where $\lambda = A/S$ is the area ratio, $C_{f0} = \tau_{w0}/\tfrac{1}{2}\rho U_{F0}^2$ is the natural friction coefficient defined based on the 'undisturbed' farm-layer wind speed $U_{F0}$, and $\beta = U_F/U_{F0}$ is the ratio indicating how much the farm-layer wind speed decreases from the natural state. Equation (5) is important as it describes the relationship between $\alpha$ and $\beta$ for a given set of $\lambda$, $C_{f0}$ and the ratio of wall shear stresses, $\langle \tau_w \rangle/\tau_{w0}$.

---

[1] Strictly speaking, the classical actuator disc theory is valid only for the case with uniform inflow. For the case with sheared inflow, the theory may still be used to estimate the thrust and power of an ideal turbine by replacing the 'upstream' wind speed with a corrected 'reference' wind speed (such as $U_{T0}$ defined above) but this is only approximately valid. In order for the theory to be strictly valid for the sheared inflow case, *the average of the square (for thrust) or cube (for power) of the upstream velocity of the air that is eventually passing through the turbine swept area* needs to be used in the normalisation [3].

*2.3. Modelling the wall shear stress ratio*
The only remaining issue here is how to quantify the wall shear stress ratio $\langle \tau_w \rangle/\tau_{w0}$. In reality, this stress ratio is expected to depend on the array configuration and design of the turbines as well as their operating conditions. However, since the primary aim of this study is to predict a practical upper limit to the farm efficiency, here we model this wall shear stress ratio simply as

$$\frac{\langle \tau_w \rangle}{\tau_{w0}} = \beta^\gamma = \left(\frac{U_F}{U_{F0}}\right)^\gamma \tag{6}$$

where the value of the exponent $\gamma$ is assumed to be close to but less than 2. This assumption is based on the conjecture that the 'effective' friction coefficient, defined as $C_f^* = \langle \tau_w \rangle/\tfrac{1}{2}\rho U_F^2$, would not vary significantly but may somewhat increase (as turbines tend to increase the turbulence intensity within the farm layer) from the natural friction coefficient $C_{f0} = \tau_{w0}/\tfrac{1}{2}\rho U_{F0}^2$. Note that $\beta = U_F/U_{F0} \leq 1$ and therefore $\gamma \leq 2$ is required to satisfy $C_f^* \geq C_{f0}$. If we employ this model, (5) can be rewritten as

$$1 - \beta^\gamma = \lambda \cdot \frac{1}{C_{f0}} \cdot \beta^2 \cdot 4\alpha(1-\alpha) \tag{7}$$

from which we can easily obtain the relationship between $\alpha$ and $\beta$ for a given set of $\lambda$, $C_{f0}$ and $\gamma$. The importance of the wall shear stress ratio will be discussed further in Section 4.

*2.4. Calculating the performance coefficients*
Finally, considering that the power extracted by an ideal turbine is $TU_T$ (and recalling that this $T$ is a good approximation to the 'maximum' turbine thrust that could be reached by changing the array configuration), we can predict the power coefficient of an ideal turbine in an ideal large wind farm as

$$C_P \equiv \frac{\text{Power}}{\tfrac{1}{2}\rho U_{F0}^3 A} = \frac{TU_T}{\tfrac{1}{2}\rho U_{F0}^3 A} = \frac{U_F^2 U_T}{U_{F0}^3} \cdot 4\alpha(1-\alpha) = \beta^3 \cdot 4\alpha^2(1-\alpha) \tag{8}$$

and hence we can calculate an upper limit to the value of $C_P$ (for a given set of $\lambda$, $C_{f0}$ and $\gamma$) by using the relationship between $\alpha$ and $\beta$ obtained from (7). Specifically, we can find an optimal value of $\alpha$ (which may vary between 0.5 and 1) to maximise $C_P$ in (8). Note that this power coefficient has been defined using the cube of the 'undisturbed' farm-layer wind speed, $U_{F0}^3$, which is identical to $U_{T0}^3$ because of (2). Here we can also calculate the 'local' power coefficient (defined using the cube of the actual farm-layer wind speed, $U_F^3$, instead of $U_{F0}^3$) as

$$C_P^* \equiv \frac{\text{Power}}{\tfrac{1}{2}\rho U_F^3 A} = \frac{TU_T}{\tfrac{1}{2}\rho U_F^3 A} = 4\alpha^2(1-\alpha) \tag{9}$$

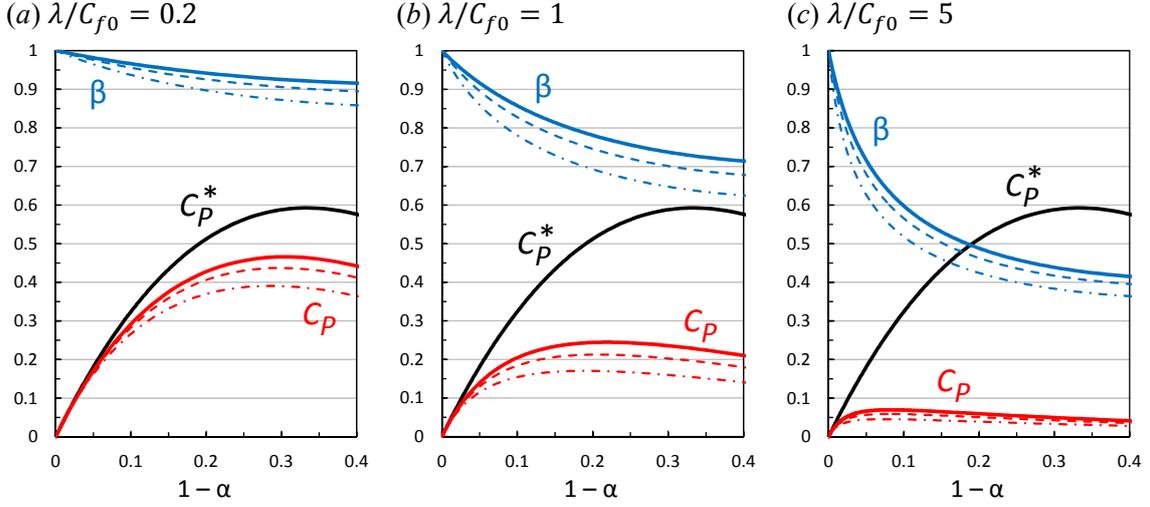

**Figure 1.** Example solutions of the two-scale momentum theory for ideal very large wind farms (solid lines for $\gamma = 2$; dashed lines for $\gamma = 1.5$; dash-dot lines for $\gamma = 1$).

Note that $C_P = \beta^3 C_P^*$ is always satisfied by definition. In addition to these power coefficients, we can also calculate a normalised power density of the wind farm, $\eta$, as follows:

$$\eta \equiv \frac{\text{Power}}{\tau_{w0} U_{F0} S} = \frac{\frac{1}{2}\rho U_{F0}^3 A \cdot C_P}{\tau_{w0} U_{F0} S} = \lambda \cdot \frac{1}{C_{f0}} \cdot C_P \quad (10)$$

Again by definition, this relationship between $C_P$ and $\eta$ is always satisfied. Hence the main point here is to obtain $C_P$ from (7) and (8); once $C_P$ has been obtained, $\eta$ can be obtained immediately.

Similarly to the power coefficients $C_P$ and $C_P^*$, we can also predict the thrust coefficient $C_T$ and the 'local' thrust coefficient $C_T^*$ of an ideal turbine in an ideal large wind farm as follows:

$$C_T \equiv \frac{T}{\frac{1}{2}\rho U_{F0}^2 A} = \beta^2 \cdot 4\alpha(1-\alpha) \quad (11)$$

$$C_T^* \equiv \frac{T}{\frac{1}{2}\rho U_F^2 A} = 4\alpha(1-\alpha) \quad (12)$$

Note that $C_T = \beta^2 C_T^*$ is always satisfied by definition.

### 2.5. Example solutions

Some examples of the solution of the above equations are presented below to demonstrate how this new theory will help us predict an upper limit to the performance of a very large wind farm. Figure 1 shows the variations of $C_P$, $C_P^*$ and $\beta (= U_F/U_{F0})$ plotted against $\alpha (= U_T/U_F)$ for three examples of ideal wind farms with the parameter $\lambda/C_{f0} = 0.2$, 1 and 5, respectively. When $\lambda/C_{f0}$ is small, e.g. when only a small number of turbines are installed in a given farm site, the farm-layer-averaged wind speed $U_F$ does not reduce significantly from its natural value $U_{F0}$ (i.e. $\beta$ remains close to 1) and therefore the power coefficient $C_P$ does not reduce significantly from the local power coefficient $C_P^*$. As $\lambda/C_{f0}$ increases, e.g. as the number of turbines installed in a given farm site increases, $\beta$ tends to decrease more significantly and so does $C_P$. It should be noted, however, that the normalised power density $\eta$ (of an ideal wind farm) increases with $\lambda/C_{f0}$. For example, if we compare two ideal wind farms with $\lambda/C_{f0} = 1$ and 0.2, the value of $C_P$ for $\lambda/C_{f0} = 1$ (at the optimal operating condition, i.e. optimal set of $\alpha$ and $\beta$) is roughly half of that for $\lambda/C_{f0} = 0.2$, meaning that the power per unit rotor swept area in the former farm is roughly half of that in the latter farm, but the power density of the former farm will be roughly 2.5 times larger than that of the latter farm (since $\eta = C_P \times \lambda/C_{f0}$).

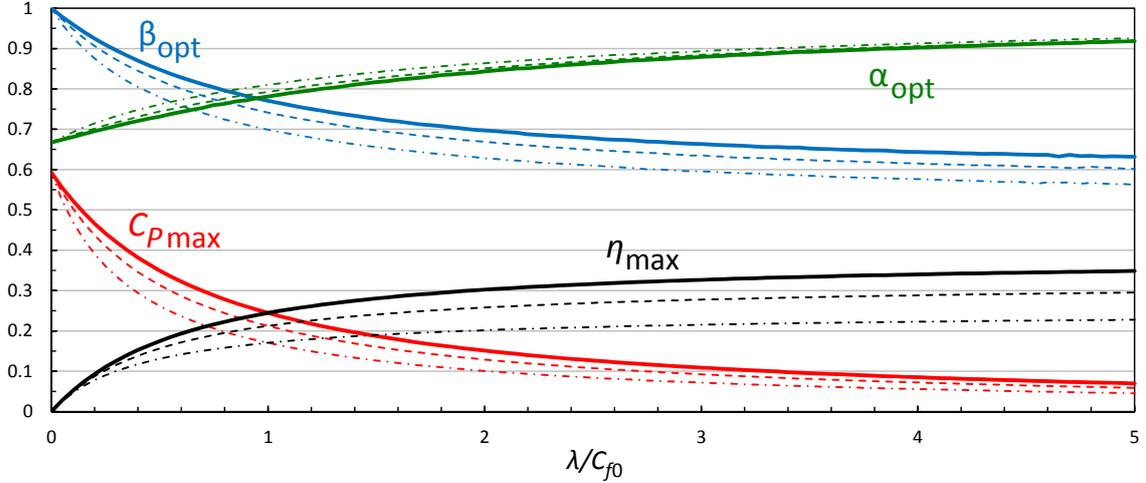

**Figure 2.** The maximum power coefficient $C_{P\mathrm{max}}$, maximum 'normalised' power density $\eta_{\mathrm{max}}$ and optimal operating conditions, $\alpha_{\mathrm{opt}}$ and $\beta_{\mathrm{opt}}$, of ideal very large wind farms, plotted against the farm parameter $\lambda/C_{f0}$ (solid lines for $\gamma = 2$; dashed lines for $\gamma = 1.5$; dash-dot lines for $\gamma = 1$).

The example solutions presented in figure 1 also show that the influence of the exponent $\gamma$ used in the modelling of the wall shear stress ratio (6) on the prediction of $C_P$ is relatively small compared to the effect of $\lambda/C_{f0}$. As noted earlier in Section 2.3, in reality the value of $\gamma$ would also depend on the turbine design and operating condition as well as the array configuration but is expected to be close to and less than 2. Therefore it seems reasonable to argue that a practical upper limit to the performance of a very large wind farm depends primarily on the farm parameter $\lambda/C_{f0}$.

Of particular interest here is how the maximum values of $C_P$ and $\eta$ (that are reached by optimising the set of $\alpha$ and $\beta$) change with the farm parameter $\lambda/C_{f0}$. Figure 2 shows the maximum values of $C_P$ and $\eta$, together with the optimal values of $\alpha$ and $\beta$, plotted against $\lambda/C_{f0}$. It can be seen that the new theory goes back to the classical actuator disc theory (i.e. $C_{P\mathrm{max}}$ approaches to the Betz limit, $16/27 \approx 0.593$, and $\alpha_{\mathrm{opt}}$ approaches to 2/3) when $\lambda/C_{f0}$ approaches to zero. Note that this corresponds to the situation where the rotor swept area $A$ is negligibly small compared to the site area for each turbine $S$; therefore $\beta = 1$ and $\eta = 0$ are always satisfied at $\lambda/C_{f0} = 0$. As $\lambda/C_{f0}$ increases, $\beta_{\mathrm{opt}}$ decreases and hence $C_{P\mathrm{max}}$ also decreases, but $\eta_{\mathrm{max}}$ increases (as already mentioned above). It should also be noted that $\alpha_{\mathrm{opt}}$ increases (i.e. the optimal local axial induction factor reduces) as $\lambda/C_{f0}$ increases.

Another interesting result of the new theory demonstrated in figure 2 is that the maximum power density $\eta_{\mathrm{max}}$ approaches asymptotically to its upper limit as the farm parameter $\lambda/C_{f0}$ increases to a very large value. However, special care must be taken when interpreting these theoretical results for large $\lambda/C_{f0}$ since, when $\lambda/C_{f0}$ is very large, it may no longer be appropriate to use (4) to estimate the 'maximum' turbine thrust $T$, i.e. the classical actuator disc theory may no longer provide a good approximation to the maximum turbine thrust in a very large wind farm; this point will be discussed further in Section 4 (following the results of the numerical analysis presented in Section 3).

## 3. Numerical analysis

A key assumption used in the new theoretical approach described above is that the classical actuator disc theory can provide a good approximation to the maximum possible turbine thrust (for a given local axial induction) not only for a single isolated turbine but also for a turbine in a very large wind farm. To validate this assumption numerically, a series of three-dimensional (3D) incompressible Reynolds-averaged Navier-Stokes (RANS) simulations of an ABL flow over a doubly periodic array of actuator discs are performed for two different types of array configurations, namely 'aligned' and 'displaced' configurations (figure 3). The streamwise turbine spacing $L_x$ is fixed at $6D$ in this study, where $D$ is the turbine rotor (or disc) diameter, whereas the lateral turbine spacing $L_y$ and lateral displacement $\Delta_y$ are varied, as summarised in Table 1.

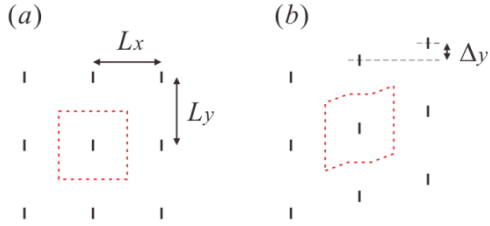

**Figure 3.** Schematic of periodic turbine array configurations: (*a*) aligned and (*b*) displaced. The red dashed line shows the computational domain.

**Table 1.** Summary of array configurations.

| Config. | $L_x/D$ | $L_y/D$ | $\Delta_y/D$ | $\lambda = A/S$ |
|---|---|---|---|---|
| Aligned | 6 | 1.5 | 0 | 0.0873 |
| Aligned | 6 | 3 | 0 | 0.0436 |
| Aligned | 6 | 6 | 0 | 0.0218 |
| Displaced | 6 | 3 | 1.5 | 0.0436 |
| Displaced | 6 | 6 | 1.5 | 0.0218 |
| Displaced | 6 | 9 | 1.5 | 0.0145 |

The simulations performed in this study are similar to those reported in [6] for a single lateral row of actuator discs (modelled as porous discs). However, there are two major differences: (i) only one disc is simulated in the computational domain with periodic boundary conditions applied not only in the lateral (*y*) but also in the streamwise (*x*) directions; and (ii) turbulent viscosity value is reduced in the vicinity of the edge of the porous disc so as to suppress changes in momentum balance due to the (otherwise undesirably strong) mixing around the disc edge. Further details are described below.

*3.1. Computational domain and flow conditions*
The height of the computational domain (or the thickness of the ABL) is $25D$, being identical to the previous study [6]. The streamwise length of the domain is fixed at $6D$ (since $L_x/D = 6$ for all cases) whilst the lateral width depends on $L_y/D$ (see Table 1). Also, for the 'displaced' cases, the domain is laterally skewed depending on $\Delta_y/D$, as shown in figure 3. The disc is located at the horizontal centre of the domain ($x = y = 0$) and near the bottom wall ($z = 0$). The vertical gap between the bottom wall and the disc is $0.5D$ and hence the disc centre is located at $(x, y, z) = (0, 0, D)$.

The top boundary of the domain is treated as a symmetry boundary, whereas the bottom boundary is treated as a smooth wall. For each simulation, the mass flow through the domain is fixed such that the cross-sectionally averaged velocity $U_{avg} = 10$m/s (instead of using a fixed streamwise pressure gradient across the domain as assumed in the theory; this point will be discussed later in Section 4). Also, for the sake of convenience, the air density and viscosity are considered to be $\rho = 1.2$kg/m$^3$ and $\mu = 1.8 \times 10^{-5}$kg/m-s, respectively, and the disc diameter $D = 100$m (resulting in a nominal Reynolds number, Re $= \rho U_{avg} D/\mu$, of about 67 million), all following the previous study [6]. Hence the 'natural' or 'undisturbed' ABL profile obtained in this numerical study (i.e. the profile obtained when the disc resistance is zero) is the same as the fully-developed sheared inflow profile employed in the previous study [6] (see figure 3 in [6]).

*3.2. Computational methods*
All simulations are performed using a commercial CFD solver 'ANSYS FLUENT 15' together with its User Defined Functions (UDF) module for modifications. The solver is based on a finite volume method, solving numerically the 3D incompressible RANS equations with the Reynolds stress terms modelled using the standard $k$-$\varepsilon$ model of Launder and Spalding [7] (with the standard wall functions applied to the bottom boundary). The numerical method used is nominally second-order accurate in space (second-order upwind for momentum, $k$ and $\varepsilon$). All simulations are steady-state simulations but a very large number (about a million) of iterations are performed for each case to confirm that all key variables, especially $U_T$ and $U_F$, converge sufficiently (with a typical uncertainty of ~0.1%).

Similarly to [6, 8, 9], the actuator disc is modelled as a stationary permeable disc (or porous disc) with a parameter (momentum loss factor) $K$ to change the disc resistance. Specifically, the impact of the disc on the (Reynolds-averaged) flow is modelled as a loss of momentum in the streamwise ($x$) direction; this momentum loss is 'locally' calculated on the disc surface (per unit disc area) as

$$M_x = K \cdot \tfrac{1}{2}\rho U_d^2 \tag{13}$$

where $U_d$ is the local (rather than disc-averaged) streamwise velocity through the disc. Since the disc-averaged thrust can be calculated as $T = \int M_x \, dA = K \cdot \tfrac{1}{2}\rho \int U_d^2 \, dA$, we obtain

$$C_T^* \equiv \frac{T}{\tfrac{1}{2}\rho U_F^2 A} = K \frac{\int U_d^2 \, dA}{U_F^2 A} \quad \text{and} \quad K = \frac{T}{\tfrac{1}{2}\rho \int U_d^2 \, dA} \tag{14a,b}$$

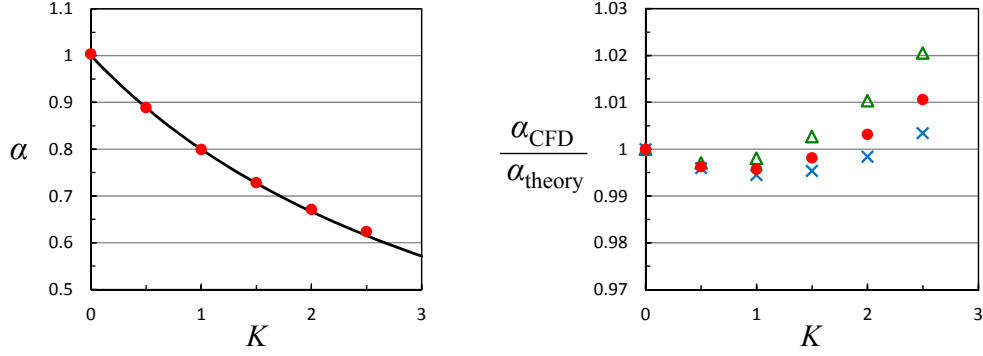

**Figure 4.** Comparison of the normalised wind speed through the disc, $\alpha$, for an isolated disc (Line: classical actuator disc theory; Symbols: CFD predictions using a porous disc model with local suppression of turbulent viscosity ($\triangle$ for $l_x = D$; $\bullet$ for $l_x = 2D$; $\times$ for $l_x = 4D$)).

Note that $K = 2$ is a theoretically optimal value for an isolated disc (corresponding to the Betz limit, regardless of whether the inflow is uniform or sheared [3]), although this optimal value can be larger than 2 for closely-arrayed discs due to the so-called 'local blockage effect' [6].

A problem in using the above porous disc model in 3D RANS simulations to represent an actuator disc (or an ideal 'Betz' rotor), however, is that turbulent mixing near the edge of the disc results in substantial 'errors'. For example, for a single disc placed near the bottom of a very wide domain with sheared inflow (corresponding to the '1 disc' case in [6]), the values of $\alpha$ and $C_T^*$ obtained using the above porous disc model at $K = 2$ are about 5.5% and 11.4% over-predicted, respectively, compared with the classical actuator disc theory, mainly because the strong mixing between the 'core' flow and 'bypass' flow near the disc edge causes additional changes in the streamwise momentum balance [8]. In order to reduce these 'errors' (but without impairing significantly the ability of RANS simulations to predict the mixing characteristics of the ABL), the turbulent viscosity $\mu_t$ is suppressed locally near the disc edge as $\mu_t = \rho f_\mu C_\mu k^2/\varepsilon$ (where $C_\mu = 0.09$, following the standard $k$-$\varepsilon$ model [7]) with

$$f_\mu = 1 \quad (\text{for } |x| \geq l_x + R), \qquad f_\mu = 1 - f_r \quad (\text{for } |x| \leq l_x), \tag{15a,b}$$

$$f_\mu = 1 - f_r \cdot \tfrac{1}{2}\left(1 + \cos\frac{|x|-l_x}{R}\pi\right) \quad (\text{for } l_x \leq |x| \leq l_x + R) \tag{15c}$$

where

$$f_r = 0 \quad (\text{for } |r - R| \geq R), \qquad f_r = \tfrac{1}{2}\left(1 + \cos\frac{|r-R|}{R}\pi\right) \quad (\text{for } |r - R| \leq R) \tag{16a,b}$$

$R = 0.5D$ is the disc radius and $r = \sqrt{y^2 + (z - D)^2}$ is a radial coordinate (with its origin at the disc centre). Figure 4 compares the normalised average wind speed through the disc $\alpha$ obtained using the porous disc model with the above modification to $\mu_t$ (with the localisation parameter $l_x = D$, $2D$ and $4D$) with that predicted by the classical actuator disc theory (again for the '1 disc' case with sheared inflow in [6]). It can be seen that now the agreement with the theory is very good, although the value of $\alpha$ still tends to be slightly over-predicted at a large $K$ value if the parameter $l_x$ is too small. In this study we use $l_x = 2D$, with which the 'error' in the value of $\alpha$ obtained is less than 0.5% for $K \leq 2$.

**Table 2.** Summary of computational results ($\alpha = U_T/U_F$, $\beta = U_F/U_{F0}$).

| Case | K=1 α | β | $C_T^*$ | $C_P^*$ | K=2 α | β | $C_T^*$ | $C_P^*$ |
|---|---|---|---|---|---|---|---|---|
| Aligned ($L_y/D$ = 1.5) | **0.747** | 0.699 | **0.559** | **0.418** | **0.629** | 0.642 | **0.793** | **0.500** |
| Aligned ($L_y/D$ = 3) | 0.703 | 0.754 | 0.494 | 0.347 | 0.583 | 0.702 | 0.680 | 0.397 |
| Aligned ($L_y/D$ = 6) | 0.619 | **0.816** | 0.383 | 0.237 | 0.507 | **0.780** | 0.514 | 0.261 |
| Displaced ($L_y/D$ = 3) | 0.777 | 0.762 | 0.604 | 0.470 | 0.649 | 0.706 | 0.842 | 0.546 |
| Displaced ($L_y/D$ = 6) | 0.792 | 0.821 | 0.628 | 0.497 | 0.663 | 0.767 | 0.879 | 0.583 |
| Displaced ($L_y/D$ = 9) | **0.800** | **0.850** | **0.639** | **0.511** | **0.670** | **0.799** | **0.899** | **0.603** |

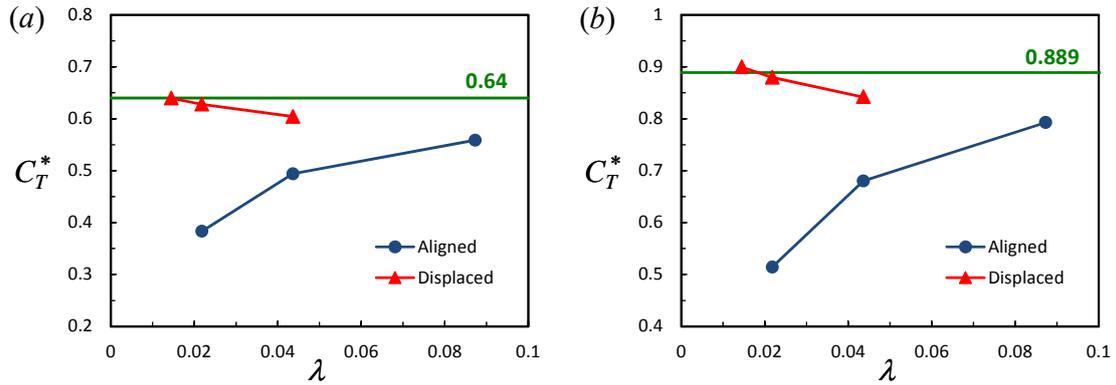

**Figure 5.** Local thrust coefficient $C_T^*$ for 'aligned' and 'displaced' cases: (*a*) $K$ = 1; (*b*) $K$ = 2.

*3.3. Computational grids*
Multi-block structured grids composed of hexahedral cells are used in this study. A 2D (*y-z*) mesh is created first and then extruded to the third (*x*) direction to form a 3D mesh. The 2D mesh around the disc is essentially the same as the 'normal resolution' mesh validated and used in the previous study [6] (shown in figure 4 in [6]). For the third direction, however, a fine uniform grid spacing of 0.02*D* (which is nearly the same as the finest grid spacing used is the 'normal resolution' 2D mesh) is used across the entire domain of 6*D* long in this study, unlike with the previous study [6] using a fine grid spacing only near the disc. The reason for using such a fine streamwise grid spacing in this study is to minimise discretisation errors in the wake region (especially for the 'displaced' cases, for which the mesh is moderately skewed at $-3 \leq x/D \leq -1$ and $1 \leq x/D \leq 3$). The total number of cells in the final 3D mesh varies between 820,800 (for $L_y/D$ = 1.5) and 3,772,800 (for $L_y/D$ = 9).

*3.4. Results*
The 'undisturbed' flow profile (obtained with $K$ = 0) was analysed first to determine the farm height $H_F$ (defined earlier in Section 2.2). The wind speed averaged over the disk area was calculated to be 8.69m/s, whereas that averaged over $0 \leq z \leq H_F$ was calculated to be 8.68m/s for $H_F$ = 2.5*D*. Hence this $H_F$ = 2.5*D* is considered as the farm height in this CFD study.

Table 2 summarises all computational results obtained for the 'aligned' and 'displaced' cases with $K$ = 1 and 2. Also, figure 5 shows the values of $C_T^*$ plotted against the area ratio $\lambda$ and compares them with the classical actuator disc theory, which yields a thrust coefficient of 16/25 = 0.64 for $K$ = 1 and 8/9 ≈ 0.889 for $K$ = 2. A clear difference in the variation of $C_T^*$ can be seen between the 'aligned' and 'displaced' cases, reflecting the importance of array configuration in wind farm design. Among the six different array configurations tested, the 'displaced' array with $L_y/D$ = 9 (or the smallest $\lambda$) produces the highest $C_T^*$ for both $K$ = 1 and 2. Of particular importance here is that this highest $C_T^*$ value does not exceed (at $K = 1$) or only slightly exceeds (at $K = 2$) the value predicted by the classical actuator disc theory. This suggests that the key assumption made earlier in Section 2.2 is approximately valid, although further investigations into other array configurations are required to fully confirm this.

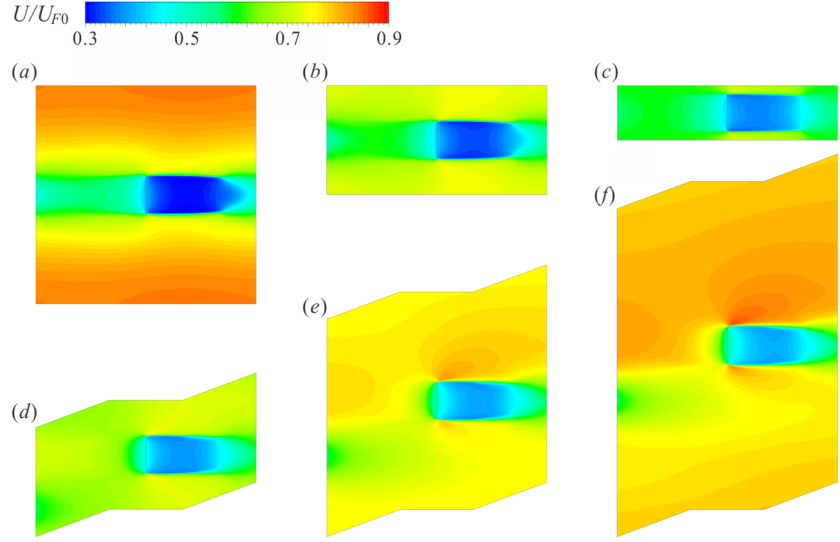

**Figure 6.** Normalised streamwise velocity contours at the disc centre height ($z = D$) for six different array configurations at $K = 2$: (*a*) aligned, $L_y/D = 6$; (*b*) aligned, $L_y/D = 3$; (*c*) aligned, $L_y/D = 1.5$; (*d*) displaced, $L_y/D = 3$; (*e*) displaced, $L_y/D = 6$; (*f*) displaced, $L_y/D = 9$.

Figure 6 shows contours of a normalised streamwise velocity, $U/U_{F0}$, at $z = D$ for the six different array configurations with $K = 2$. It can be observed that, for the 'aligned' cases, the velocity upstream of the disc tends to increase as we reduce the lateral spacing $L_y$, whereas for the 'displaced' cases, the velocity upstream of the disc decreases with $L_y$ (i.e. the benefit of displacing discs decreases).

## 4. Discussion and conclusions

In this paper a new theoretical approach has been proposed to predict an upper limit to the efficiency of a very large wind farm. The virtue of the new theory is that it helps us estimate an approximation to the highest possible farm efficiency (for a given farm site) without knowing any details of the actual flow profile in the wind farm *a priori*. As described in detail in Section 2 and highlighted in figure 2, the new theory suggests that the efficiency of ideal turbines in an ideal very large wind farm should depend primarily on $\lambda/C_{f0}$, which is a parameter we can easily obtain before constructing a farm.

It is worth noting that the new 'two-scale-coupled' momentum theory proposed here is somewhat analogous to the tidal farm model proposed by Vennell [10]. A clear difference between the wind and tidal cases, however, is that, for the wind case, the farm layer height $H_F$ needs to be somehow defined in order to couple the actuator disc theory with the larger-scale momentum balance. The definition of $H_F$ introduced in this study may, at the first glance, appear to be rather artificial, but is probably the most convenient one for the theory to be concise. By employing this definition of $H_F$, the majority of difficulties in theoretical modelling of very large wind farms are aggregated into only two key issues: (i) whether the classical actuator disc theory can give a good approximation to the 'maximum' local thrust coefficient $C_T^*$, and (ii) what is the actual value of the exponent $\gamma$ in (6). It should be noted that, in reality, the value of $\gamma$ (or alternatively the average wall shear stress $\langle \tau_w \rangle$) may also depend on the array configuration as well as the turbine design used. Therefore, for a given $\lambda/C_{f0}$, a particular array configuration that gives the highest $C_T^*$ and/or $C_P^*$ (for a given $\alpha$) may not necessarily yield the highest $C_P$ (because a different array configuration might give a larger $\gamma$ (or smaller $\langle \tau_w \rangle$) resulting in a large enough $\beta$ for $C_P = \beta^3 C_P^*$ to be higher). This essentially means that when we try optimising our turbine design and/or array configuration for a very large wind farm, we should aim not only to maximise the local power coefficient $C_P^*$ but also to minimise $\langle\tau_\mathrm{w}\rangle$. However, for the purpose of predicting an upper limit, the present theory with $\gamma = 2$ could be used, as already explained in Section 2.3.

Another interesting result of the present theory is that, for $\gamma = 2$, the maximum normalised power density $\eta_\mathrm{max}$ approaches asymptotically to an upper limit value of 0.3849 (with $\alpha$ and $\beta$ approaching to 1 and 0.5774, respectively) as $\lambda/C_{f0}$ increases to a very large value. This upper limit value agrees with the one derived earlier in [11]. However, as can be seen from the CFD results for the 'displaced' arrays presented in Section 3, in reality it becomes more and more difficult to maintain the values of $C_T^*$ and $C_P^*$ as high as those predicted by the classical actuator disc theory as the area ratio $\lambda$ increases (because it becomes more difficult to avoid negative effects of wake interaction). Hence it seems that the present theory gives a 'conservative' upper limit when $\lambda/C_{f0}$ is very large.

Before concluding the paper, some remarks should be made also on the numerical study presented in Section 3. It should be stressed that this numerical study was designed to assess the validity of the key assumption used in the new theory (i.e. estimating the 'maximum' $C_T^*$ values approximately by using the classical actuator disc theory) and not the validity of the new theory as a whole. To fully validate the new theory, we also need to investigate the effect of different array configurations on the average wall shear stress $\langle\tau_\mathrm{w}\rangle$ or the exponent $\gamma$, even though the value of $\gamma$ is expected to be close to and less than 2 as explained in Section 2.3. This was not investigated in the present numerical study because RANS simulations with the standard wall functions are not capable of predicting changes in the wall shear stress accurately. This also explains why a fixed mass flow condition was employed in the present numerical study instead of a fixed pressure gradient across the domain. A higher-fidelity numerical simulation, such as large-eddy simulation (LES), would be helpful on this point.

Finally, it should be noted that this two-scale momentum theory could, of course, be extended or modified in the future. For example, the farm-scale (or ABL-scale) momentum balance considered in Section 2.1 could be modified to include the effect of the Coriolis force and other physical effects. It may also be possible to employ some corrections to the actuator disc theory to consider, e.g. the local blockage effect [6] (this may help explain the reason why $C_T^*$ can slightly exceed the value predicted by the classical actuator disc theory). Nevertheless, the theory presented in this paper does seem to explain the most fundamental part of this interesting problem and is expected to remain useful as a baseline theory for the aerodynamics of very large wind farms.